\documentclass[12pt,preprint]{aastex}

\shorttitle{MOND Kinematic Self-Consistency}
\shortauthors{C.W. Stubbs and A. Garg}
\begin{document}
\title{ Testing the Self-Consistency of MOND With
Three Dimensional Galaxy Kinematics}
\author{Christopher W. Stubbs and Arti Garg} 

\affil{Department of Physics \\
and \\
Harvard-Smithsonian Center for Astrophysics \\
Harvard University, Cambridge MA USA}
\email{cstubbs@fas.harvard.edu, artigarg@fas.harvard.edu}

\begin{abstract}

We propose a technique to test the idea that non-standard dynamics,
rather than dark matter halos, might be responsible for the observed
rotation curves of spiral galaxies.  In the absence of non-luminous
matter, a galactic disk's rotational velocity and its vertical velocity
dispersion can be used jointly to test the self-consistency of the
galaxy's dynamics.  A specific illustrative example, using 
recent measurements of the disk kinematics of M33, shows this  
to be a promising approach to assess the viability of Modified Newtonian Dynamics (MOND).
 
\end{abstract}

\keywords{Dark Matter---galaxies:kinematics and dynamics---galaxies:
individual(\objectname {M33})}

\clearpage
\section{INTRODUCTION}

Understanding the structure and kinematics
of spiral galaxies, in particular explaining their rotation curves at
large galactic radii, remains one of the pressing open questions in astrophysics.
Optical observations of galactic rotation curves find that rather than falling off as one
would expect from galaxy models where the mass traces the observed
light, the rotational velocities remain constant at large
radii.  These findings are further borne out
by radio observations of the
21 cm line, from HI gas in the outer parts of the
galactic disk. An overview of rotation curves is provided in \cite{Rubin01}, \cite{Persic96} and 
\cite{Salucci01}. 

There is strong evidence from the CMB data for considerable amounts of 
non-baryonic dark matter on the cosmological scale \citep{WMAP03}. While is it 
enticing to imagine that the dark matter problems on the galactic and cosmological 
scales have a common resolution, this need not necessarily be the case. Our focus
in this paper will be on the galactic dark matter problem as manifested in the 
ubiquitous observation of flat rotation curves. 

A number of ideas have been put forth to account for flat rotation
curves.  All require some form of new physics.  These ideas include:

\begin{enumerate}

\item{} New particles: Missing mass in the form of galactic dark matter, most likely non-baryonic \citep{Alcock00}.

\item{} New interactions: Non-gravitational long-range couplings might exist, or gravitational physics might be subject to revisions over large distances.

\item{} New dynamics: scenarios such as Modified Newtonian Dynamics (MOND) in which 
gravity from visible matter is the only force acting but the system's response takes on new aspects.
 
 \end{enumerate}

The community consensus at present prefers the dark matter hypothesis, but galactic dark matter has thus far evaded all attempts to detect it. We should strive to test, whenever possible, alternatives to the dark matter scenario. 

\subsection{The MOND Approach to the Rotation Curve Puzzle: Novel Dynamics}

  Motivated by the observed spiral galaxy rotation curves,
  \citet{Milgrom83} proposed a modification of the dynamics of
  non-relativistic matter.  This modified behavior, termed
  MOND for Modification of Newtonian Dynamics, is conjectured to
  arise only in the regime of low accelerations. MOND is a proposed
  modification to an object's acceleration under an applied force,
  such that $a=g/\mu(x)$ where $g$ is the acceleration expected under
  Newtonian physics, and $x=a/a_0$ depends upon the MOND acceleration
  scale $a_0 \sim 1.2 \times 10^{-10} m/s^2$, with $\mu(x \gg 1) \simeq
  1, \mu(x \ll 1) \simeq x$.  A commonly adopted form is $\mu(x) =
  {{x}/{\sqrt{1+x^2}}}$.  In general a gravitating system's behavior
  under MOND can be described by taking the Newtonian description and
  replacing $G$, the coupling constant, by $G/\mu$, with the
  understanding that the dynamics is being altered rather than the
  nature of the gravitational interaction.
  
In this scenario the mass of a galaxy resides in the ordinary
 astronomical components that we can detect by their emission or
 absorption of electromagnetic radiation, and the galaxy's light
 distribution traces out its mass distribution.  A review of MOND as
 an alternative to dark matter is presented in \citet{Sanders02}.

  In the MOND model the response of a test particle to an applied
  force depends upon the magnitude of its absolute acceleration
  relative to a preferred frame, taken to be the local rest frame of
  the microwave background. ``Overacceleration'' at low values of $x$
  then produces the observed rotation curves of spiral galaxies.  MOND
  thereby eliminates the need for dark matter, at the expense of novel
  dynamics at low accelerations.  MOND does a remarkably good job of
  fitting the rotation curves of galaxies across a wide range of
  surface brightness, with $a_0$ as the single free parameter
  \citep{Sanders02}.

  The MOND idea was recently placed on a more formal footing \citep{Beckenstein04}, 
  but our approach will be cast in the original phenomenological framework, 
  in terms of $\mu(x)$. From this standpoint, the formulation described
  above suggests an observational test for the self-consistency of MOND.
  Because $\mu(x)$ depends on the (scalar) magnitude of a particle's
  total acceleration, comparing the vertical and rotational dynamics of test
  particles in the disk of a spiral galaxy provides a means to test for self-consistency.
  This paper proposes a framework for carrying out such a test and
  illustrates the technique with recent data from M33 \citep{Ciardullo04}.
  
Other recent efforts to investigate the viability of the MOND
hypothesis using kinematics include using galaxy clusters
\citep{Silk05} and globular clusters of stars \citep{Baum05}.  Our
approach differs in that, as discussed in the following section, we
are checking the self-consistency of MOND rather than comparing 
the observed kinematics to a prediction.
  
  \section{USING 3-d DISK KINEMATICS TO TEST SELF-CONSISTENCY}

  We will adopt the common model for a galactic disk as a mass distribution with a 
  volume matter density $\rho(r,z)$ given by 
\begin{equation}
\rho(r,z)= \rho_0 e^{-r/R_0} sech^2(z/z_0)
\end{equation}
  \noindent
  where $R_0$ and $z_0$ are characteristic scale lengths in the radial
  and vertical directions.  In the MOND scenario, the rotational speed
  of a thin galaxy with a mass structure described by equation (1)
  obeys (adapting the expression for Newtonian physics from
  \citet{Pad02})

  \noindent
  \begin{equation}
  v^2(r) = a(r)~ r = {g(r) \over \mu_r(r)}  r = {4 \pi G \Sigma_0 R_0 \over \mu_r(r)}
   y^2 [I_0(y)K_0(y) - I_1(y)K_1(y)]
  \end{equation} 
  \noindent
  where $a$ and $g$ are the MOND and Newtonian accelerations respectively, 
  $G$ is Newton's gravitational constant, 
  $R_0$ is the disk scale length, 
  $\Sigma_0$ is the surface mass density at $r=0$, 
  $y=r/2R_0$ is a dimensionless radius variable, $I_n(y)$ and
  $K_n(y)$ are $n^{th}$ order modified Bessel functions, 
  and $\mu_r$ gives rise to an object's 
  modified response in the radial direction.  

  The vertical kinematics of the objects in the disk are described by                                                                                    
  $\sigma_z^2(r) = 2 \pi G \Sigma(r) z_0/\mu_z(r)$, where $\Sigma(r)$ is the
  local surface mass density, 
  $z_0$ is the vertical scale height, and $\mu_z$ 
  accounts for MONDian behavior in the vertical direction.
  Given these two expressions and the exponential radial form for $\Sigma(r)$ given in equation (1),
  we can construct the dimensionless ``Consistency Parameter" ratio $CP(r)=\mu_r(r)/
  \mu_z(r)$. 
  
  {\em Regardless of the local dynamical law that describes a star's  
  response to feeble forces,  $CP(r)$ should be 
  unity, at all radii.   This is true even if the system makes a transition from the Newtonian to the 
  MOND  regime.} 

  The expressions given above allow us to write $CP(r)$ as   
\begin{equation}    
    CP(r) = {\frac {\mu_r(r)} { \mu_z(r)}} = 
    2{\frac{R_0}{z_0(y)}} {\frac{\sigma^2_z(y)}{v^2_c(y)}}
    [y^2 e^{2y} (I_0(y)K_0(y) - I_1(y)K_1(y))]. 
 \end{equation}

    For a disk-only system, the combination of the observables on the right side should 
    equal to unity at all radii in both the
    Newtonian (where $\mu$=1) and MONDian (where $\mu <$ 1) regimes.  $CP(r)$
    can be understood as a parameter testing whether the tracer material
    is behaving in a self-consistent fashion.  Note that this expression
    is independent of both the mass-to-light ratio of the disk material
    and of the central surface density of the disk.
      
      Ideally, one would obtain both face-on and edge-on observations of a
      single galaxy.  This would then allow the measurement of the rotation curve, of the
      vertical velocity dispersion, and of the scale lengths $R_0$ and $z_0$. In practice, this is of course not possible. 
       
       There is, nevertheless, a realistic possibility of measuring the
       {\it radial dependence} of $CP(r)$, using kinematic information alone. A nearly face-on bulgeless spiral galaxy
       would provide a powerful testbed. The apparent (projected) rotation curve
       would be suppressed by an unknown $sin (i) $ factor, where $i$ is the
       inclination angle, giving an observed $v_{obs}=v_c sin(i)$. This would simply 
       rescale $CP$ by an overall
       multiplicative factor, so that while still radius--independent it will differ from 
       unity by $1/sin^2(i)$.  Given that typical rotational velocities are a
       few 100 km/s while velocity dispersions are tens of km/s, a tenfold
       suppression of the rotation curve is quite tolerable. This corresponds
       to using systems with inclination angles as small as a few degrees.  The advantage
       of using a face-on galaxy is that the line-of-sight velocity
       dispersion is a clean measure of vertical velocity dispersion,
       uncontaminated by other components of the velocity ellipsoid within
       the galactic disk. 

       The main observational challenge is obtaining high
       signal-to-noise measurements and then extracting both the circular velocity field and the
       vertical velocity dispersion. 
       This is a tractable problem. The projected circular velocity produces a Doppler-shifted 
       centroid of a spectral feature. The velocity dispersion can be determined from either the 
       broadening of spectral lines, or by the width of the velocity distribution of a set of individual 
       resolved objects. 

       We emphasize the fact that as long as the disk scale height $z_0$ is independent of $r$, 
       and the self-supporting stellar disk dominates $\rho(r,z)$, only velocity data are needed  to test
       for variation in $CP(r)$.  

       By selecting nearly face-on galaxies for this test, we lose the ability to 
       measure their vertical scale height and must instead appeal to a statistical argument that
       invokes measurements of edge-on analogous systems. Measurements of the light distribution of edge-on galaxies in the near infrared, using images from the 2Mass survey  \citep{Bizyaev02} indicate that for typical galaxies the vertical scale height {\em is} 
       independent of galactic radius, with typical values of $z_0/R_0$ varying between 0.1 and 0.4. 
     We will therefore adopt the working hypothesis that 
       $z_0$ in equation (3) is independent of $r$. 

	 \section{AN ILLUSTRATIVE KINEMATIC SELF-CONSISTENCY TEST USING M33}

	 There are a few instances where a galaxy disk's vertical velocity
	 dispersion has been measured (e.g.~\citet{Bottema93}).  A recent data set on the
	 kinematics of M33 \citep{Ciardullo04} provides an interesting test case. These
	 authors obtained line-of-sight velocity data on 140 planetary
	 nebulae in M33. This archival data set provides an opportunity for a concrete
	 example of the $CP$-violation test outlined above. 

	 \subsection{The Kinematic Properties of M33}

	 This local group galaxy is inclined at 56 degrees to the plane of the sky. 
	 Various determinations (using multiple techniques) yield a distance modulus of 
	 24.8 $\pm$ 0.1 {\it mag}. The radial scale lengths for light are  
	 $R_0^V=$ 2.5 kpc and $R_0^K=$ 1.56 kpc in the V and K bands, respectively.   
	 This implies
	 (taking the 2Mass-derived typical values for $R_0/z_0$) a likely vertical
	 scale height $z_0$ in M33 of a few hundred parsecs. 

	 The M33 inclination angle of 56 degrees is not optimal for
	 our purposes, but Ciardullo et al (2004) provide their best
	 estimates of M33's rotational velocity and vertical velocity
	 dispersion as a function of galactocentric distance. Their
	 results are presented in Table 1.  The M33 circular velocity
	 at $r$=10 kpc implies, in the MOND scenario, that objects at
	 that radius should experience a threefold increase in their
	 radial (and hence vertical) acceleration, relative to the
	 Newtonian value.  The typical acceleration component normal
	 to the disk is $a_{z}\sim \sigma^2_z/z_0$, well into in the
	 MOND regime. It is therefore the radial component that
	 determines the kinematics of the objects in the disk.  An
	 extended rotation curve for M33 is presented in
	 \citet{Corbelli00}.

\begin{deluxetable}{rrrrr}
\tabletypesize{\scriptsize}
\tablecaption{Kinematic Properties of M33, from \citet{Ciardullo04}.
The
	 first column lists galactocentric distance, the second and third the
	 rotational velocity and vertical rms velocity dispersion, and the fourth column
	 shows the inferred value of $\mu_r$, the radial MOND dynamical parameter, based
	 on the measured circular velocity and radial distance. The circular velocity has 
	 been corrected for projection effects and represents the best estimate for the actual rotation 
	 curve of the galaxy.}

\tablewidth{0pt}
\tablehead{
\colhead{ R(kpc)} & \colhead{$v_c~(km/s)$} & \colhead{$\sigma_z$ (km/s)} &\colhead{~~$\mu_{radial}$~~}
}
\startdata
0.5 &   40 &    21~~ &    0.66  \\ 
	 1 &     55 &    18.7 &  0.64  \\ 
	 2 &     80 &    17~~ &    0.66   \\ 
	 3 &     90 &    14~~ &    0.60  \\ 
	 4 &     98 &    12.5 &  0.55 \\ 
	 5 &     100 &   10.5 &  0.49 \\ 
	 6 &     105&    10~~ &    0.45  \\ 
	 7 &     106&    8~~ &     0.41 \\ 
	 8 &     107&    7.5 &   0.37  \\ 
	 9 &     108&    5.5 &   0.34  \\ 
	 10 &    109 &   5~~ &     0.31 \\
\enddata
\label{tab:Table1}	  
\end{deluxetable}

	The observational data in Table 1, in conjunction with equation (3), provide us with the opportunity to map out the radial dependence of the consistency parameter, $CP$, across the face of M33. 

\subsection{A Mass-Traces-Light MOND Consistency Analysis}

In the MONDian view, the ordinary astronomical inventory of M33, plus
novel dynamics, produce the observed rotation curve. The light
distribution across M33 should then trace the galaxy's mass
distribution. We have evaluated the radial dependence of $CP$ using
values of $R_0$ obtained from visible and near-IR wavelengths.

The measurements of M33's kinematics from Table 1 were used in
conjunction with equation (3) to determine $CP(r)$, for different
values of the structural parameters $R_0$ and $z_0$. One choice we
made was to set $R_0=$ 1.56 kpc and $z_0=$ 0.4 kpc, corresponding to
roughly the midpoint of the 2Mass aspect ratio distribution. The
resulting $CP(r)$ values are shown in Figure 1. Since for the M33 data
$R_0$ is better constrained than $z_0$ we also explored values of
$z_0$ that produced the $CP(r)$ curve closer to unity, by fitting for
$z_0$ while minimizing the sum $\Sigma(1-CP(r))^2$ for the radii
listed in Table 1. With $R_0$ fixed at 1.56 and 2.5 kpc, the best-fit
values of $z_0$ were 0.23 and 0.079 kpc, respectively, corresponding
to values of $z_0/R_0$ of 0.14 and 0.03. The best-fit value of $z_0$
for the larger disk scale length is only 80 pc, and corresponds an
unreasonably thin disk. The corresponding $CP(r)$ curves for these
cases are also shown in Figure 1.

The curves in Figure 1 indicate that, as shown in equation (3), the value of $R_0$ determines the {\it shape} of the $CP(r)$ curve, and the $z_0$ parameter only provides an overall multiplicative scaling that can be adjusted to drive the average $CP$ value towards unity. 

In this mass-traces-light analysis, the $CP$ parameter changes by a factor of 3--10 (depending upon the fit used) between the inner and outer regions of the M33 disk. In this illustrative example the $CP(r)$ behavior appears inconsistent with the MOND scenario. The sense of the discrepancy, with $CP$ values less than one, corresponds to $\mu_z>\mu_r$, so that matter appears to be overaccelerating in the radial direction more than in the vertical direction. 

\begin{figure}
\plotone{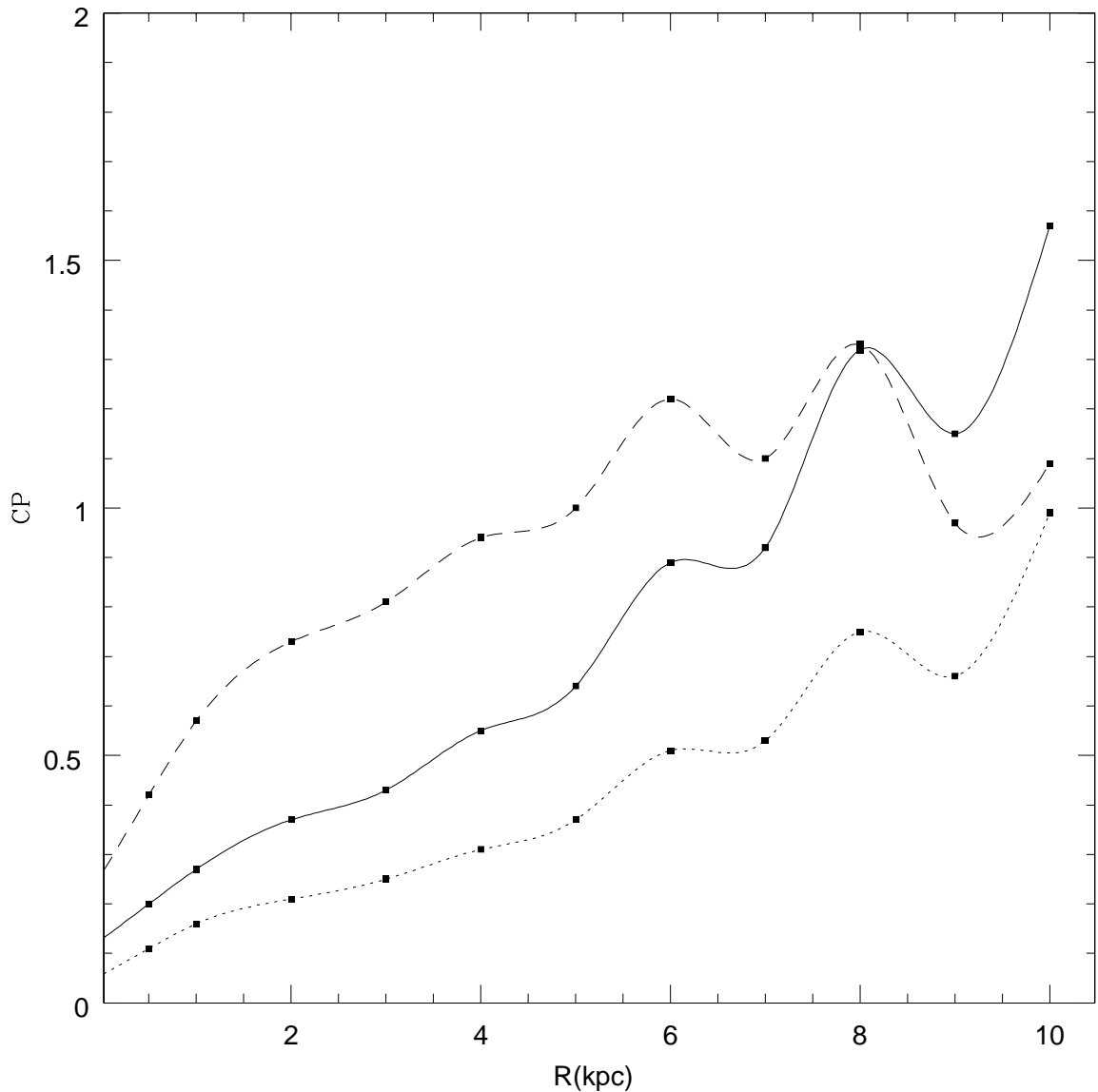}
\caption{Values of the dimensionless Consistency Parameter, $CP$, in M33 vs. galactic radius, for light-traces-mass scenarios. The plot shows the radial dependence of $CP=\mu_r/\mu_z$, the ratio of the MOND overacceleration parameter in the radial and vertical directions.  The curves show $CP(r)$ for galactic structural parameters  (in kpc) of ($R_0, z_0$) equal to 
(1.56, 0.23)=solid, (1.56, 0.4) = dotted, and (2.5, 0.079) = dashed. A self-consistent MOND galaxy would have $CP=1$ at all radii.} 
\end{figure}
 
 \subsection{Relaxing the Mass-Traces-Light Constraint: Allowing a Wider Range of $R_0$ and $z_0$}
 
The exponential scale length of the M33 disk emission depends upon the passband used to measure the surface brightness, ranging from 1.56 kpc in the K band to 2.5 kpc in the V band. Stepping back from the light-traces-mass approach, is interesting to explore what combinations of $R_0$ and $z_0$ would produce a $CP(r)$ closest to unity. 

Allowing both $R_0$ and $z_0$ as free parameters, with no constraints and with no assumption about the galaxy's mass-to-light ratio, the values that best match  $CP=1$ are $R_0=5.4$ kpc and $z_0=0.035$ kpc.  This corresponds to a remarkably thin disk, with a vertical scale height of only 35 pc. 

The $CP(r)$ profile from this more general fit is shown in Figure 2. This provides a better fit to the kinematic observations, but $CP$ still varies by over a factor of two across the face of M33. Also, the value of  $z_0/R_0=0.006$ is over an order of magnitude less than typical aspect ratios. 

Although we can achieve improved MOND-inspired fits to the kinematic data, these models imply strong radial and vertical gradients in the galaxy's mass-to-light ratio. This is at variance with the elegant what-you-see-is-all-there-is MOND scenario. 
 
\begin{figure}
\plotone{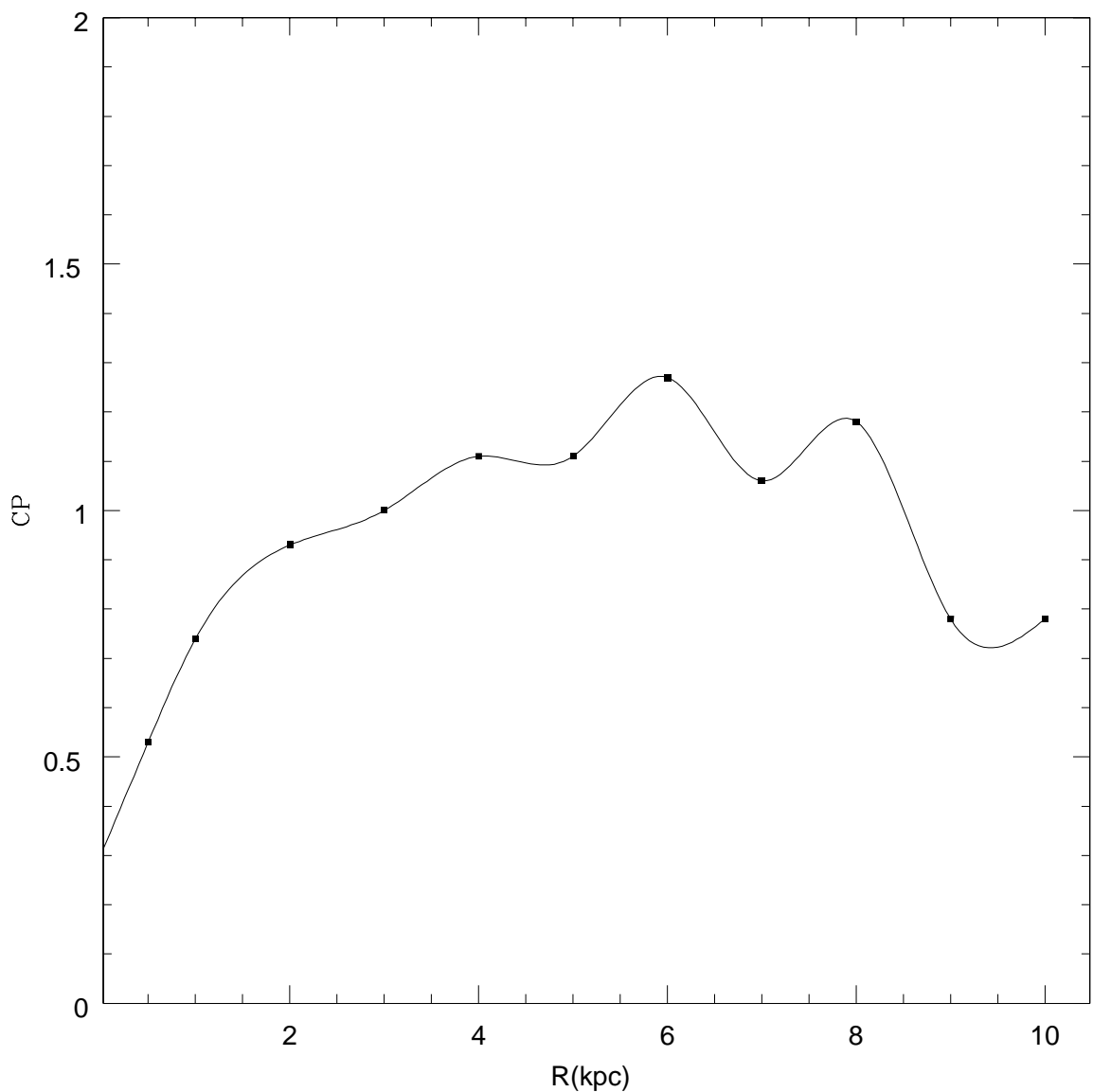}
\caption{Consistency Parameter $CP$ vs. radius after relaxing the light-traces-mass constraint. The solid line corresponds to $(R_0, z_0)=$ (5.4, 0.035) kpc, the best fit to $CP=1$ when both are allowed as unconstrained free parameters. This kinematic fit is improved over those shown in Figure 1, but still exhibits significant variation with radius. Also the resulting aspect ratio is at variance with that seen in other similar galaxies.}
\end{figure}

\section{DISCUSSION}

Our objective is to propose a general technique for testing the self-consistency of MOND, using the existing M33 data as an illustrative example. The vertical and circular motions of a galaxy can be jointly used for this test. Potential weaknesses in the argument presented above include i) the assertion that the vertical scale height of galaxies is radius-independent, ii) modeling the galaxy with the form shown in equation (1), and iii) the implicit assertion that either objects overaccelerate, or they don't. The first issue can be addressed with better observations and more statistics, and the second by a more comprehensive treatment of the system's kinematics. 

The third concern, namely the isotropy of MONDian dynamics, is an
interesting issue. If MONDian behavior arises from a modification of
inertia \citep{Milgrom05}, then this scalar quantity will determine an
object's response to {\it any} applied force, and it will exhibit the
same modified dynamics in all directions. On the other hand one might
imagine that MOND only applies component by component, with a modified
response only to those forces that would give rise to accelerations
below the $a_0$ threshold. This could produce a difference in the
radial and vertical dynamics and could perhaps account for a ratio of
$\mu_{radial}/\mu_{vertical}$ that differs from unity. In this
circumstance however a terrestrial Cavendish experiment conducted at
the North or South pole should see differing effective values of $G$ in
different regimes of $\mu$.

Sensible next steps to obtaining observations that are optimally
suited to the test we propose include 1) assessing the relative merits
of planetary nebulae vs. integrated starlight as probes of vertical
velocity dispersion, 2) selecting a favorable list of target galaxies,
and 3) carrying out a set of appropriate observations. It is sensible
to include, as a control, examples of high surface brightness disk
galaxies which should have their inner regions in the Newtonian
disk-dominated regime where $\mu$ =1, to verify that $CP$ is constant
and equal to unity for these systems. H$\alpha$ and 21 cm observations
of the velocity field might also contribute to this technique.

\acknowledgments

We are grateful to J. Beckenstein, G. Bothun, J. Dalcanton, and K. Cook for
interesting conversations about MOND as an alternative to dark
matter. J. Battat, A. Miceli, D. Sherman and the thoughtful students in Harvard's Fall 2005 freshman seminar on the Hidden Universe provided important encouragement.  We thank Harvard University and the Department of Energy Office of Science for their support.

\end{document}